\documentclass[preprint,12pt,authoryear]{elsarticle}

\usepackage{amssymb}
\usepackage{amsthm}

\graphicspath{ {./figures/} }

\usepackage{amsmath}
\usepackage{bm} 
\usepackage{mathbbol} 
\usepackage[]{textcomp} 
\usepackage{enumitem}
\usepackage{csquotes}
\usepackage{caption}
\usepackage{setspace} 
\usepackage[left=1.5cm, right=1.5cm]{geometry} 

\usepackage{blindtext}
\usepackage{soul}
\usepackage[normalem]{ulem}

\usepackage[dvipsnames]{xcolor}

\usepackage{longtable,tabu} 
\usepackage[section]{placeins}

\usepackage{color,soul}
\setstcolor{red}

\journal{Journal of the Mechanical Behavior of Biomedical Materials}

\begin{document}
	
	
	\begin{frontmatter}
		
		
		\title{Fluid-solid interaction in the rate-dependent failure of brain tissue and biomimicking gels}
		
		\author[a,b]{M. Terzano}\corref{cor1}
		\author[a]{A. Spagnoli}
		\author[b]{D. Dini}
		\author[c,d]{A. E. Forte}
		
		\address[a]{Department of Engineering and Architecture, University of Parma, Parco Area delle Scienze 181/A, 43124 Parma, Italy}
		\address[b]{Department of Mechanical Engineering, Imperial College London, Exhibition Road, London SW7 2AZ, UK}
		\address[c]{DEIB, Politecnico di Milano, Via Ponzio, 34/5 - 20133 Milano, Italy}
		\address[d]{School of Engineering and Applied Sciences, Harvard University, Cambridge, Massachusetts, USA}
		
		
		\begin{abstract}
			
			Brain tissue is a heterogeneous material, constituted by a soft matrix filled with cerebrospinal fluid. {The interactions between, and the complexity of each of these components } are responsible for {the non-linear} rate-dependent behaviour {that characterizes what is one of the most complex tissue in nature}. {Here, we investigate the influence of the cutting rate on the fracture properties of brain, through wire cutting experiments}. {We also present a model for the rate-dependent behaviour of fracture propagation in soft materials, which comprises} the effects of fluid interaction through a poro-hyperelastic {formulation}. The method is {developed in the framework of finite strain continuum mechanics,} implemented in a commercial finite element code, and applied to the {case} of an edge-crack remotely loaded by {a} controlled displacement. Experimental and numerical results {both show } a toughening effect with {increasing} rates, which is linked to the energy dissipated {by the} fluid-solid interactions in the process zone ahead of the crack.
			
		\end{abstract}
		
		
		
		\cortext[cor1]{Corresponding author. Tel.: +39-0521-905927; fax: +39-0521-905924;\\
			e-mail: spagnoli@unipr.it.}
		
	\end{frontmatter}
	
	
	
	\section{Introduction}

	Brain tissue is arguably one of the most complex, delicate and heterogeneous tissues of the human body. Its structure is characterised by two main constituents: the grey matter, which contains the nerve cell bodies, and the white matter, with a large proportion of myelinated axons \citep{Budday2020a}. By a mechanical point of view, neural tissues are among the softest of all internal organs \citep{Guimaraes2020}, receiving protection from the skull and isolation from external actions by the cerebrospinal fluid. A large proportion of this fluid is free to move by diffusion and consolidation within the tissue's solid network; as a result, the brain behaves as a soft sponge: its microstructure, albeit highly inhomogeneous, presents small pores that are saturated by fluid \citep{Forte2017}. Diffusion has a fundamental importance for the brain function, delivering vital nutrients to the neural cells and playing an essential role in therapies based on drug delivery \citep{Nicholson2001}. Besides, the motion of fluid within the solid network causes volumetric shrinking and triggers consolidation effects \citep{Franceschini2006a}, which can explain various phenomena, including the onset and evolution of hydrocephalus and the brain shift during surgeries \citep{Stastna1999, Forte2018}. The interaction between interstitial fluid and solid matrix provides a source of energy dissipation \citep{Mak1986}, which results in time-dependent behaviour{,} frequently observed during mechanical testing \citep{Jin2013, Forte2017}. In addition, a further source of dissipation is related to viscoelasticity, caused by intracellular interactions between cytoplasm, nucleus and the cell membrane \citep{Budday2017}.
	
	Mechanical models of the brain tissue at the continuum scale are usually formulated in the framework of finite strain mechanics, addressing the nonlinear elastic, time-dependent and {hysteretical} behaviour \citep{DeRooij2016}. The biphasic nature {of the tissue} can be captured by models derived from the classical theory of consolidation in soil mechanics \citep{Biot1941, Franceschini2006a, Forte2017}, eventually coupled with large deformations \citep{Simon1992, Garcia2009, Hosseini-Farid2020}. An equivalent description has been developed in the context of mixture theories \citep{Mow1980}, leading to the formulation of a consistent framework for soft porous media \citep{Ehlers2015, Comellas2020}. Time-dependent behaviour due to viscous effects has been described by generalised Maxwell models \citep{Forte2017, Qian2018} or more refined descriptions elaborated in the finite strain theory \citep{Budday2017a, Haldar2018}. 
	{However, }with respect to tissue failure, our understanding is {considerably} more limited. {It is known that} the brain tissue, as most internal organs, does not carry significant mechanical loads; {nevertheless}, traumatic injuries expose the tissue to damage and fracture \citep{ElSayed2008a}. Furthermore, {the tissue can be perforated with catheters, needles and probes, during} minimally invasive surgeries and regenerative therapeutics \citep{Casanova2014, Ashammakhi2019, Terzano2020}. Due to its high heterogeneity, failure {properties in} the brain tissue {are} region dependent. In the white matter{, which is characterized by} fibrous axonal structures, failure occurs by tearing of fibres when {the tissue is} loaded above a certain threshold \citep{Budday2020a}. At the microscale, axonal injury involves a viscoelastic mechanism with stretching and sliding of microtubules, depending on the entity of the deformation \citep{Cloots2011, Ahmadzadeh2014}. While the contribution of fluid-solid interaction in terms of the tissue mechanical behaviour is widely recognised, the role of fluid diffusion during failure has {not been investigated}. Furthermore, the flow of fluid in the brain tissue is not homogeneous \citep{Zhan2019, Jamal2020}. Due to the different microstructure, white matter is far more permeable than grey matter, which instead presents densely connected networks that can entrap the fluid phase \citep{Budday2020a}. Such a difference might explain the enhanced rate-dependence of white matter during compression and tensile tests, because of a faster fluid drainage \citep{Budday2020a}. However, the characteristic time of fluid draining depends on the size of the perturbed region, which also makes this contribution dependent on testing conditions \citep{Wang2012a}. {Therefore, there is a need for investigating how rate-dependency and fluid-structure interactions affect fracture propagation in brain tissue.}
	
	When a porous network is swollen by fluid, mechanical and hydraulic responses are coupled: forces and deformations change the pressure of interstitial fluid, while pressure gradients drive fluid flow, resulting in mechanical deformation \citep{Arroyo2017}. During fracture, the flow of fluid inside the crack-tip zone might affect the surface energy required for crack initiation and propagation. Despite scarce information in the context of biological tissues, illuminating evidences come from experimental work on failure of hydrogels. {As an example,} studies on reversible gels suggest that the fracture energy can be increased by the drainage of fluid in the crack-tip zone \citep{Baumberger2006a, Naassaoui2018}. In addition, polymeric gels swollen with fluid and subjected to subcritical loading, i.e. such that the elastic strain energy is not {sufficient} to cause instant failure, delay {their failure} {because of the increase of available energy fracture created by the fluid drainage}  \citep{Wang2012b}. 
	
	In this work, our aim is to shed light on the rate-dependent fracture process in the brain tissue caused by fluid draining. Firstly, we present the results of fracture tests performed on porcine brain samples. {To this aim we use the wire cutting protocol, which} is a well established method to measure the fracture properties of soft materials, including viscous foodstuff and gels \citep{Goh2005, Baldi2012, Forte2015a}. A computational model is then developed in the framework of finite strain continuum mechanics, representing the large strain behaviour and fluid interaction through a poro-hyperelastic model \citep{Simon1992}. The numerical analyses are focused on the process of crack propagation, which in {the case of} wire cutting develops after the initial stages of indentation and tissue rupture \citep{Terzano2020a}. Through a simplified model of the fracture process in dissipative materials \citep{Zhao2014}, we are able to consider the energy dissipated by fluid-structure interaction as a function of the loading rate. {Finally, we provide } a comparison {is provided} with the poroelastic behaviour of a biomimicking gel that was previously characterised by \citet{Forte2015a}.
	
	
	\section{Materials and methods}
	\label{sec:methods}
	
	\subsection{Wire cutting tests}
	\label{subsec:exp}
	
\begin{figure*}[htb]
		\includegraphics[scale=1]{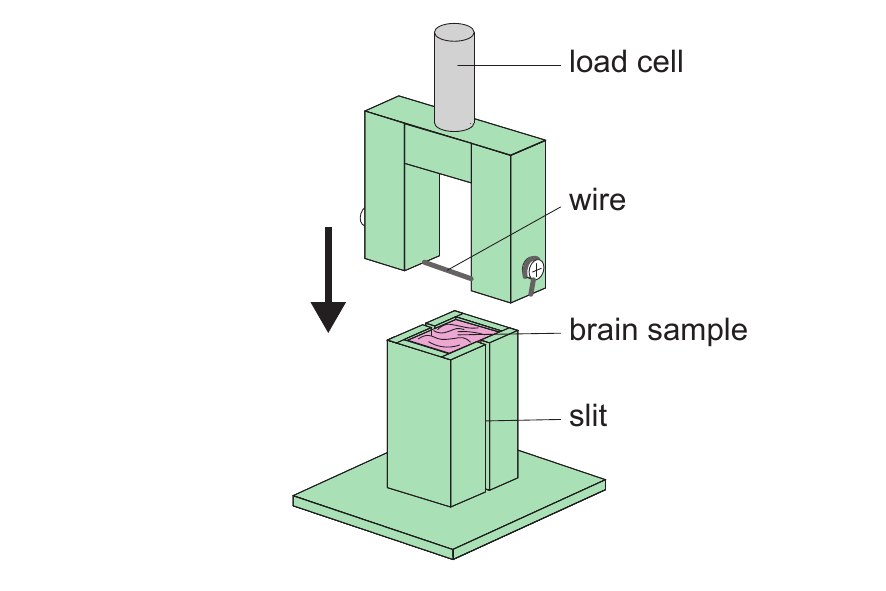}
		\caption{Wire cutting testing schematic. \textcolor{red}{1-col figure}}
		\label{fig:wire_schematic}
	\end{figure*}

	When measuring the fracture toughness of soft materials, traditional techniques based on stress intensity factors cannot be employed, since failure occurs when a large portion of the material is well beyond the limit of small strain elasticity. Toughness is hereby defined as the total amount of energy absorption during deformation until fracture occurs \citep{Huang2019}. Wire cutting tests were preferred with respect to other available methods (such as, for instance, edge-notched tensile tests \citep{Long2016}) because of the issues related to the extreme softness of the brain tissue, the effect of self-weight and the impossibility of realising proper clamping. Porcine brain tissue samples were prepared, removing the cerebellum and separating the two hemispheres; each hemisphere was then positioned in the sample container with the frontal lobe facing upwards. The specimen would slowly shift under gravity and approximately occupy a square prism of length 30 mm, width 30 mm and height 50mm. Steel wires of diameter  $ d_w=0.05, 0.16, 0.25, 0.5\mathrm{mm} $  were inserted with an insertion speed of $ v=5\mathrm{mm \,s^{-1}} $, and the test with $ d_w=0.16\mathrm{mm} $ was repeated with $ v=0.5\mathrm{mm \,s^{-1}} $ and $ v=50\mathrm{mm \,s^{-1}} $. All tests were performed with a Biomomentum Mach-1\texttrademark mechanical testing system using a 1.5 N single-axis load cell, in a conditioned room at 19 °C temperature \citep{Forte2016}. {A schematic of the experimental setup is shown in Fig. \ref{fig:wire_schematic}}.

	\subsection{Model of poroelastic fracture}
	\label{subsec:model}
	
	A model is proposed to account for the rate-dependence observed during wire cutting tests on the porcine brain samples, which can be extended to similar materials with a soft and wet porous microstructure. It is based on the following assumptions: (i) brain tissue and the biomimicking gels are modelled as poro-hyperelastic materials; (ii) rate-dependent failure is described with a model of the fracture process based on the spatial separation of dissipative length scales and the definition of a cohesive process zone; (iii) fracture in cutting is assimilated to the propagation of a far-field loaded crack, depending on a geometric parameter (in this specific case, the wire diameter).
	
	\subsubsection{Poro-hyperelastic model}
	
	The brain tissue and the biomimicking gels are considered as biphasic materials, where a soft solid skeleton is saturated by biological fluids \citep{Franceschini2006a, Forte2017, Comellas2020}. In this section we describe a poroelastic model at large strains, with specific focus on the equations needed for its numerical implementation in a finite element (FE) code. The theory of finite deformation continuum mechanics, as presented in standard textbooks on the subject, e.g. \citep{holzapfel2000nonlinear}, is the background in which the model is developed. For the sake of consistency with an updated Lagrangian framework in which the incremental solution strategy is implemented in the commercial FE code, the field equations are referred to the current configuration. Accordingly, a material point of the biphasic medium is identified by the position vector $ \bm{x}(\bm{X},t) $, $ \bm{u}_S \equiv \bm{u}= \bm{x}-\bm{X}(\bm{x},t) $ is the displacement of this point in the porous solid phase and $ \bm{u}_F $ defines the corresponding quantity for the pore fluid \citep{Simon1992} (Fig.\ref{fig:sketch}a). We also recall the decomposition of the spatial velocity gradient $ \mathbf{l}=\nabla\dot {\bm{u}}= \mathbf{d} + \mathbf{w} $, where $ \mathbf{d}=\mathrm{sym}(\nabla\dot {\bm{u}}) $ is the symmetric rate of deformation tensor, $ \mathbf{w}=-\mathbf{w}^T $ is the anti-symmetric spin tensor and $ \dot {\bm{u}} $ is the velocity of the solid phase.\footnote{Throughout this section, $ \nabla(\bullet) $ denotes the spatial gradient while $ \nabla\cdot(\bullet) $ is used for the spatial divergence operator. Italic is used for scalars, bold italic for vectors and bold roman for tensors.}
	
	In a biphasic material, each phase in the current configuration is defined by a volume fraction $ n_\alpha=\mathrm{d}v_\alpha/\mathrm{d}v $, where $ \alpha=S,F $ corresponds, respectively, to the solid skeleton and the pore fluid. Assuming conditions of saturation, we establish the fundamental relationship $ n_F + n_S=1 $ \citep{cheng2016poroelasticity}. In the following, we denote $ n=n_F $ the porosity of the medium, which is correlated to the current void ratio through $ e=n/(1-n) $. The continuity mass equation for phase $ \alpha $ reads \citep{Ehlers1999}
	\begin{equation}\label{eq:mass1}
		\frac{\mathrm{D}}{\mathrm{D}t}	(n_\alpha \rho_\alpha) + n_\alpha \rho_\alpha\nabla \cdot \dot {\bm{u}}_\alpha=0
	\end{equation}
	where $ {\mathrm{D}(\bullet) }/{\mathrm{D}t}$ is used for the material time derivative and $ \rho_\alpha $ is the effective density of each phase. In the solid skeleton, Eq.(\ref{eq:mass1}) provides the following relationship
	\begin{equation}\label{eq:mass2}
		\frac{(1-n)}{(1-n_0)}=J_SJ^{-1}
	\end{equation}
	where $ n_0 $ is the porosity in the initial configuration, $ J $ is the volume ratio of the biphasic material and $ J_S $ is instead referred to the solid skeleton. Notice that if the matrix is assumed incompressible, we have $ J_S=1 $; however, this does not lead to macroscopic incompressibility, because volume change can occur through changes in the volume fractions. Considering an incompressible fluid phase, Eq.(\ref{eq:mass1}) can be rewritten as
	\begin{equation}\label{eq:mass3}
		\frac{\mathrm{D}}{\mathrm{D}t}	n + n \nabla \cdot \dot{\bm{u}}_F=0
	\end{equation}
	
	The strong form of the linear momentum balance for the biphasic material in quasi-static conditions is provided by \citep{Simon1992}
	\begin{equation}\label{eq:equil1}
		\nabla \cdot \bm{\sigma} + \rho \bm{b}=\bm{0}
	\end{equation}
	where $ \bm{\sigma} $ is the Cauchy stress tensor, $ \rho=(1-n)\rho_S + n\rho_F $ is the homogenised density and $ \bm{b} $ is the vector of body forces per unit mass. The corresponding weak form can be written as
	\begin{equation}\label{eq:virtual1}
		{\int_{\Omega}\bm{\sigma}:\delta\mathbf{e}\;\mathrm{d}v }- 	{\int_{\Omega}\rho\bm{b}\cdot \delta\bm{u}\;\mathrm{d}v - \int_{\partial\Omega_t}\bar{\bm{t}}\cdot\delta\bm{u}\;\mathrm{d}a}=0\quad \forall \delta\bm{u}
	\end{equation}
	where $ \delta\mathbf{e}=\mathrm{sym}\nabla\delta\bm{u} $, with $ \delta\bm{u} $ being the virtual solid displacement, and $ \bar{\bm{t}} $ is the prescribed traction vector on the boundary $ \partial\Omega_t $. In order to obtain the constitutive equations of the biphasic material in a suitable formulation for incremental Newton's type numerical methods, a rate form should be introduced \citep{holzapfel2000nonlinear}. The first term in Eq.(\ref{eq:virtual1}) represents the internal virtual work and taking its material time derivative we have
	\begin{equation}\label{eq:rate1}
		\dot{\delta W_\mathrm{int}}=\int_{\Omega}(\bm{\sigma}:\dot{\delta\mathbf{e}} + \dot{\bm{\sigma}}:\delta\mathbf{e})\;\mathrm{d}v
	\end{equation}
	where $ \dot{\delta\mathbf{e}} = - \mathrm{sym}(\nabla\delta\bm{u}\;\mathbf{l})$ \citep{holzapfel2000nonlinear}. In order to express the material rate of the Cauchy stress, we refer to the well-known concept of the effective stress in biphasic materials $ {\bm{\sigma}'}=\bm{\sigma} + p_F\mathbf{I} $ \citep{Biot1941}, where $ p_F=-1/3\,\mathrm{tr}\bm{\sigma}  $ is the scalar pore pressure. Then, the second term of the parenthesis in Eq.(\ref{eq:rate1}) can be expressed by
	\begin{equation}\label{eq:rate2}
		\dot{\bm{\sigma}}={\bm{\sigma}'}^J + \mathbf{w}\cdot\bm{\sigma} + \bm{\sigma}\cdot\mathbf{w}^T - \dot{p_F}\mathbf{I}
	\end{equation}
	where we have introduced the objective Jaumann rate of the effective Cauchy stress $ {\bm{\sigma}'}^J $. The use of this specific objective rate is motivated by the implementation of the model in the commercial software Abaqus \citep{abaqus2017}. For completeness, we recall that a discretised and linearised form of the previous equations is then needed, where the virtual displacement field is approximated with suitable interpolation functions and variations are computed with respect to the field variables of the problem \textemdash which in our case are represented by nodal displacements $ \bm{u} $ and pore pressure values $ p_F $.
	
	Finally, we introduce the constitutive assumptions for the biphasic medium. The fluid flow through the porous skeleton is characterized by Darcy's law, with an isotropic permeability tensor which remains unchanged during the deformation. In quasi-static conditions and neglecting inertia, Darcy's law correlates the rate of fluid volume to the pressure gradient
	\begin{equation}\label{eq:Darcy}
		\dot{\bm{w}} =  - \frac{\kappa}{\eta_F}{\nabla}{p_F},
	\end{equation}
	where $ \dot{\bm{w}}=n(\dot{ \bm{u}}_F-\dot {\bm{u}}) $ is the seepage velocity, representing the rate of fluid volume flowing through a unit normal area, $ \kappa $ is the intrinsic permeability and $ \eta_F $ is the fluid viscosity \citep{cheng2016poroelasticity}.
	
	The behaviour of the solid skeleton is specified by a hyperelastic isotropic strain energy function. Several studies related to brain mechanics \citep{Budday2017, Forte2017} have shown that a modified one-term Ogden model provides optimal fit to experimental data \citep{Ogden1972a}. The compressibility of the solid skeleton is implemented through the usual decomposition of the solid deformation into isochoric and volumetric parts, such that the strain energy density is provided by
	\begin{equation}\label{eq:hyperOgden}
		\Psi(\bar \lambda _i) + U(J)={\frac{{{2\mu}}}{{{\alpha}^2}}\left( {\bar \lambda _1^{{\alpha}} + \bar \lambda _2^{{\alpha}} + \bar \lambda _3^{{\alpha}} - 3} \right)} + \frac{1}{D}(J-1)^{2}
	\end{equation}
	where  $ \mu $, $ \alpha $ and  $ D=2/K_S $ are material parameters and $ \bar \lambda_i = J^{-1/3}\lambda_i $ are the modified principal stretches \citep{holzapfel2000nonlinear}. The effective Cauchy stress tensor is split into its deviatoric and volumetric components $ \bm{\sigma}'=\mathbf{s}' + p'\mathbf{I} $ \citep{Selvadurai2016}, with $ p'=\partial U/\partial J $ and  \citep{Connolly2019}
	\begin{equation}\label{eq:stress2}
		\mathbf{s}'=J^{-1}\beta_i(\bm{n}_i \otimes \bm{n}_i)
	\end{equation}
	where $ \bm{n}_i $ are the principal spatial directions and the stress coefficients are expressed by $ \beta_i =\bar \lambda _i  \partial  \Psi/\partial \bar\lambda _i  - 1/3 \bar \lambda _j  \partial  \Psi/\partial \bar\lambda _j $ (the summation rule applies to repeated indices). The last step required for the numerical implementation in the updated Lagrangian framework is to make explicit the objective rate introduced in Eq.(\ref{eq:rate2}) through a spatial fourth-order elasticity tensor, such that
	\begin{equation}\label{eq:rate3}
		{\bm{\sigma}'}^J = \mathbb{c}'_J:\mathbf{d}
	\end{equation}
	where $  \mathbb{c}'_J $ is the spatial elasticity tensor defined in terms of the Jaumann rate of the Cauchy stress \citep{Crisfield1997}. The explicit formulation for a compressible hyperelastic model in terms of the principal stretches can be found, for instance, in the recent work by \citet{Connolly2019}.
	
	\subsubsection{Rate-dependent fracture process}
	
	\begin{figure*}[htb]
		\includegraphics[scale=1]{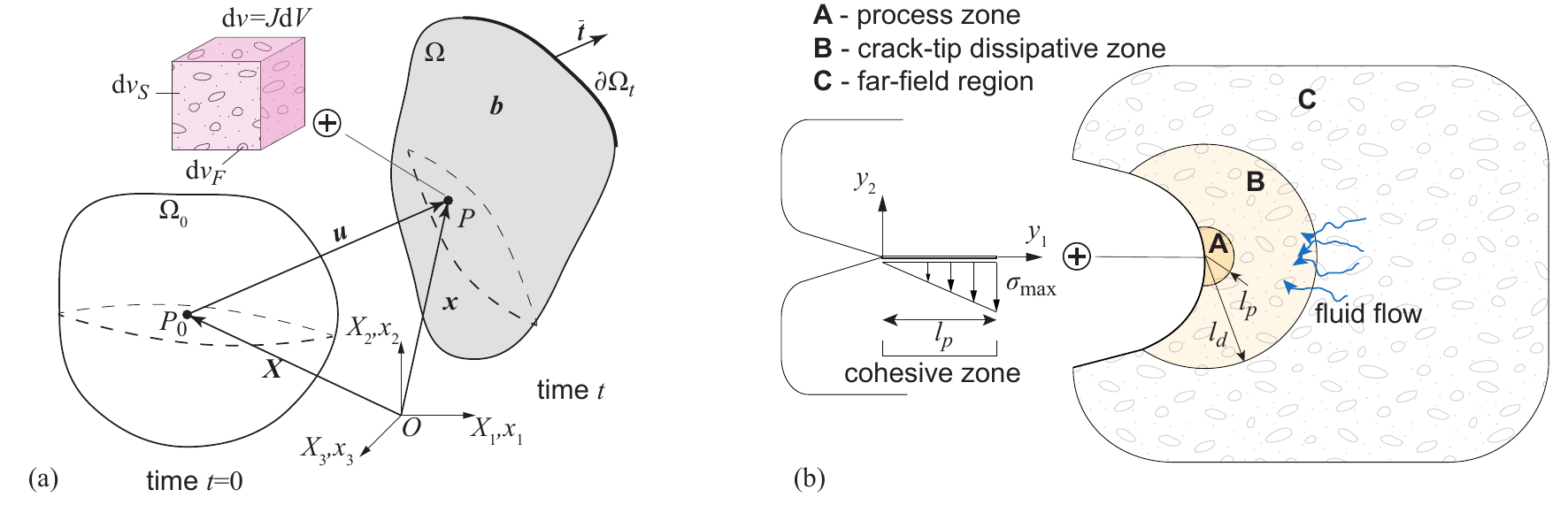}
		\caption{(a) Reference and current configurations of a biphasic continuum body. (b) Illustrative sketch of the fracture process in a poroelastic soft material. Shown in figure are (A) the process zone with radius $ l_p $, (B) the crack-tip dissipative zone with radius $ l_p $ and (C) the far-field region. The enlarged view shows the process zone schematised with a rate-independent cohesive zone model.  \textcolor{red}{2-col figure}}
		\label{fig:sketch}
	\end{figure*}

	The flow of interstitial fluid in the pores of the soft solid skeleton results in time-dependent deformation and draining of the biphasic medium, according to a relaxation time which depends on the material properties (namely, the permeability) and the length of macroscopic observation \citep{Hu2012}. The analysis of rate-dependent fracture requires that poroelastic relaxation is considered as a source of energy dissipation correlated to crack propagation \citep{Creton2016}, in which the length of observation is put in relation with some characteristic size of the fracture process.
	
	The model here proposed is based on the ideal situation illustrated in Fig. \ref{fig:sketch}b, where a propagating crack in a semi-infinite body is shown with three regions where dissipative phenomena possibly occur \citep{Long2016}. Firstly, we consider damage phenomena occurring at the molecular scale, which are condensed within the so-called process zone and account for the intrinsic toughness of the material. Within this region, energy dissipation might be affected by the loading rate and result in a toughening effect which is proportional to the velocity of crack propagation. For instance, \citet{Forte2015a} explained the rate-dependent toughening in gelatins through a mechanism of fluid draining in the pores within the process zone. At a larger scale, dissipative terms are originated from relaxation in the bulk material but become relevant to crack propagation only if they affect the crack-tip region, which we broadly define as the material affected by the vicinity of the crack. Usually, their effect is to prevent the crack driving force provided by external loading from being fully delivered to the crack \citep{Qi2018}. Finally, bulk processes in the far-field zone, which can cause macroscopic relaxation, are neglected as they do not contribute to the fracture process. As a consequence of the proposed decomposition, we assume to split the fracture energy in two terms: the intrinsic term $ \Gamma_o $ originating from the process zone, and an additional term $ \Gamma_d $ due to energy dissipation in the crack-tip region affected by propagation \citep{Zhao2014}, so that we have
	\begin{equation}\label{eq:G-diss}
		\Gamma=\Gamma_o+\Gamma_d
	\end{equation}
	
	Unlike the intrinsic toughness, $ \Gamma_d $ cannot be treated as a property of the material because it is affected by rate. With respect to the size of the crack-tip dissipative zone $ l_d $, we introduce a further hypothesis. In elastic soft materials, crack propagation is coupled with large deformations, and this is motivated by the fact that the energy cost of creating new surfaces is comparable to the elastic strain energy in the material. A length scale can be defined, known as elasto-adhesive length or length of flaw sensitive failure \citep{Creton2016, Chen2017}, which separates by several order of magnitudes most stiff solids from soft tissues. To a first approximation, it also represents the radius of a blunted propagating crack in an elastic material \citep{Creton2016}. We wish to clarify that this has nothing to do with energy dissipation but simply characterises the concept of softness by a fracture mechanics point of view. Here we assume that the size of the crack-tip dissipative zone is coincident with that of the large strain region, such that we have
	\begin{equation}\label{eq:R}
		l_d \equiv \varrho_o=\Gamma_o/E, \quad l_d \gg l_p
	\end{equation}
	where $ E $ is the initial Young's modulus of the material. In this work, it is assumed that energy dissipation is due to the drainage of fluid in the crack-tip region, that is, we are not considering viscoelasticity, while the intrinsic toughness $ \Gamma_o $ is considered rate-independent. Conceptually, this is equivalent to employ a cohesive process zone which enriches the continuum poro-hyperelastic model with a prescribed rate-independent stress-displacement relationship on the separation interface \citep{Schwalbe2012} (see the enlarged view in Fig. \ref{fig:sketch}b). In line with our assumptions, the characteristic length $ l_p $ of the cohesive region is much smaller than the size of the crack-tip dissipative region $ l_d $. Notice that, although several time-dependent cohesive models have been proposed, e.g. \citet{Noselli2016}, our approach is to consider that relaxation occurs outside the cohesive zone.
	
	\subsubsection{Fracture process in cutting}
	\label{subsubsec:cutting}
	
	Cutting involves deformation, friction and fracture. Here we focus on the steady-state, which is developed after the initial stage of contact and indentation when the external work is converted into fracture energy for crack propagation \citep{Terzano2018}. Our aim is to establish the limits under which crack propagation in cutting can be compared to propagating a crack in symmetric far-field loading conditions. Furthermore, the plain strain assumption is introduced. A schematic of the model is shown in Fig. \ref{fig:wire}a.
	
	With respect to crack propagation under remote loading, the finite size of the cutting tool adds an additional length to the fracture process in cutting. It is speculated that the tool exerts some sort of constraint on the elastic blunting of the crack, which can be limited by the fact that the crack opening displacement is determined by the tool geometry \citep{Hui2003, Zhang2019d}. In an elastic material, the tip radius of a blunted crack $ \varrho_o $ in condition of propagation is a material property and represents a characteristic length of the fracture process. It can be compared with the wire diameter $ d_w $ in order to distinguish two different scenarios \citep{Terzano2020a}:
	\begin{itemize}
		\item for $ d_w \ge 2{\varrho_o} $, crack propagation happens as an autonomous process under symmetric mode-I conditions, with a certain distance between the wire and the crack tip. The crack tip radius is determined by its natural value $ \varrho_o =\Gamma_o/E $;
		\item for $ d_w < 2{\varrho_o} $, the shape of the blunted crack is constrained by the wire, which touches the crack-tip. In this situation, the mechanism of propagation is different from that under remote loads and requires a further input of external energy. In other terms, crack propagation is energy limited.
	\end{itemize}
	
	The analyses of fracture described in this work applies to wire cutting only when the first condition is met. With this assumption, we have neglected the role of friction, which is known to affect the fracture toughness of a material \citep{Spagnoli2019a}. Furthermore, we have also considered that the criterion derived in an elastic situation is extended to the poro-hyperelastic material.
	
	
	\section{Results}
	\label{sec:results}
	
	\subsection{Experimental}
	
	\begin{figure*}[h!]
		\includegraphics[scale=1]{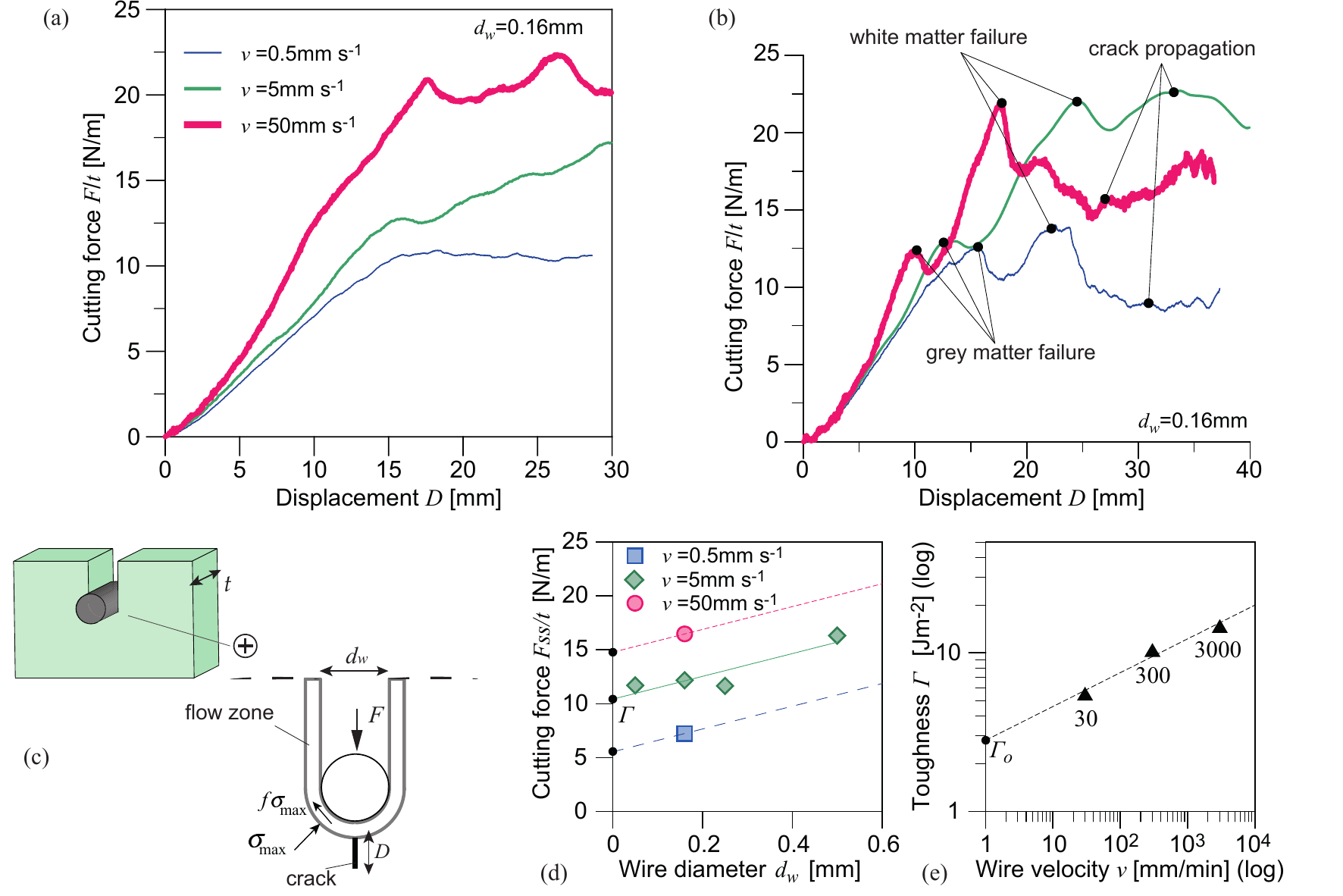}
		\caption{(a) Cutting force versus displacement for $ d_w=0.16\mathrm{mm} $. The curves show the average of various tests on porcine brain tissue, with a high dispersion (not shown here) due to the inhomogeneous structure of the samples. (b) Force-displacement from single tests at the three different insertion speeds, showing the corresponding stages of material failure. (c) Schematic of wire cutting showing the model of the rupture process proposed by \citet{Kamyab1998}. (d) Steady state force $ F_{ss}/t $ as a function of the wire diameter. The continuous linear fit is obtained for $ v=5\mathrm{mms^{-1}} $. (e) Logarithmic plot of the intrinsic toughness as a function of the insertion speed. \textcolor{red}{2-col figure}}
		\label{fig:exp}
	\end{figure*}
	
	Force-displacement curves obtained from wire cutting tests on the porcine brain tissue are illustrated in Fig.\ref{fig:exp}a, for a wire diameter $ d_w=0.16\mathrm{mm} $ and three insertion speeds (average of various tests). Following the initial indentation, in which the tissue deforms prior to fracture, the force tends to stabilise in the steady state phase of cutting \citep{Terzano2018}. Differently from the results of similar tests on gelatins \citep{Forte2015a}, the transition to the steady state is not well marked in the brain tissue, due to the extreme softness and the inhomogeneous structure of the samples. In Fig.\ref{fig:exp}b we show the force-displacement curves for single tests, where one can distinguish two peaks, corresponding to grey and white matter failure, followed by relaxation, before reaching an approximately stable trait where the wire cuts through the sample. 
	
	Wire cutting can be employed to infer the intrinsic toughness of the tissue $ \Gamma_o $. To do so, we need to remove the contribution due to energy dissipation; typically, this means performing a fracture test at very low loading rates, so that quasi-static conditions are assumed \citep{Persson2005a}. The results are here elaborated according to the model proposed by \citet{Kamyab1998}. Briefly, the steady-state cutting force $ F_{ss} $ results from the force needed to open the crack and a contribution due to the formation of a flow zone around the bottom half of the wire, as shown in Fig.(\ref{fig:exp}c). Friction produces a circumferential stress in this region but is neglected anywhere else. Then, the force per unit thickness is proportional to the wire diameter, according to
	\begin{equation}\label{eq:cutting}
		\frac{F_{ss}}{t} = \Gamma  + \left( {1 + f} \right){\sigma _\mathrm{max}}{d_w}
	\end{equation}
	where $ \sigma _\mathrm{max} $ has to be intended as a characteristic cohesive stress of the material, $ f $ is the frictional coefficient and $ t $ is the out-of-plane thickness of the sample.

	The steady-state force $ {F_{ss}}/{t} $ obtained from the cutting experiments at $ v=5\mathrm{mm\,s^{-1}} $ is plotted as a function of the wire diameter in Fig.\ref{fig:exp}d. Since a steady value cannot be easily recognised, $ F_{ss} $ is computed as the average force corresponding to the onset of crack propagation observed in the tests. A linear fit is employed to extrapolate the force to zero diameter, such that, according to Eq.(\ref{eq:cutting}), the value of the fracture toughness is obtained. However, the experimental value of the force $ {F_{ss}}/{t} $ for $ v=0.5\mathrm{mm\,s^{-1}} $ is lower, suggesting that there might be an extra contribution due to energy dissipation resulting in $ \Gamma > \Gamma_o $. Lacking complete data for lower velocities, due to the complexity of realising proper tests on the super-soft brain tissue, we then hypothesize that the same force-diameter slope applies to any insertion speed and extrapolate to $ d_w=0 $ the corresponding steady-state forces. These are shown on a logarithmic plane in Fig.(\ref{fig:exp}e) and fitted with a linear interpolating function. The intercept with the vertical axis, corresponding to a quasi-static value of the insertion speed, should provide the correct value of $ \Gamma_o $. Furthermore, the increase of toughness with speed follows a power-law, with exponent approximately equal to 0.2. Due to the uncertainty in experimental data, we might assume that a value of $ \Gamma_o $ comprised between $ 0.1-1 \mathrm{J\,m^{-2}} $ is a reasonable approximation. Indeed, this is the same order of magnitude of the toughness of biomimicking gelatins computed from wire cutting tests by \citet{Forte2015a}.
	
	\subsection{FE analyses}
	
	\subsubsection{Elastic crack blunting}
	
	\begin{figure*}[h!]
		\includegraphics[scale=1]{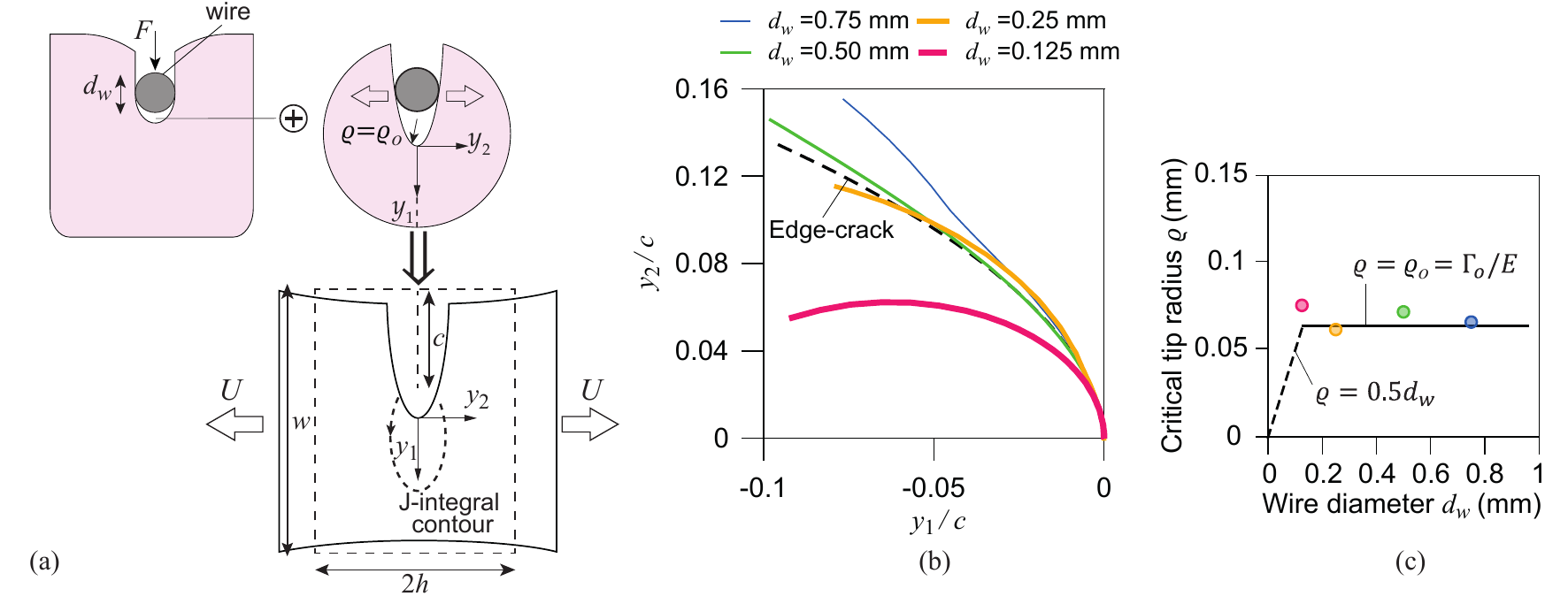}
		\caption{(a) Sketch of the plane strain geometry employed to investigate the fracture process in wire cutting. Under the assumption of $ d_w \ge 2\varrho_o $, the equivalent model of an edge crack of length $ c $, subjected to an opening remote displacement $ U $, is obtained. The local system $ y_1,y_2$ is defined with respect to the moving crack tip. (b) Deformed crack profiles in the elastic brain tissue when $ J=\Gamma_o $, for an edge-crack (dashed line) and different wire diameters $ d_w $. (c) Corresponding crack tip radius $ \varrho $ versus the wire diameter. The horizontal line is the natural crack tip radius $ \varrho_o $. \textcolor{red}{2-col figure}}
		\label{fig:wire}
	\end{figure*}
	
	Finite element analyses are employed to understand the origin of the rate-dependent fracture properties observed in experiments. In order to reduce fracture in cutting to a problem of crack propagation, we first need to verify the hypothesis presented in Sect.\ref{subsubsec:cutting}. We have modelled the steady state phase as the insertion of a rigid circular wire into an edge-cracked body of width $ w $ and height $ 2h $, with initial crack length $ c $ (Fig.\ref{fig:wire}a). Since the wire extension in the out-of-plane direction is much larger than the thickness $ t $ of the samples, a plain strain assumption can be made. Due to symmetry, only half specimen is modelled with pertinent constraints imposed to the lower edge of the body; eight-node plane strain elements are employed, with a suitable refinement around the crack tip, which is artificially blunted by taking an initial small radius of curvature. From analyses of mesh convergence, the smallest element in the crack tip region is equal to $ 10^{-5}\,h $. The sample material is purely elastic, described by the strain energy provided in Eq.(\ref{eq:hyperOgden}). In such a case, the crack driving energy is correctly provided by the \textit{J}-integral, Eq.(\ref{eq:Jint-1}), such that the onset of crack propagation occurs when $ J=\Gamma_o $. The material parameters implemented in the FE model are summarised in Table \ref{tab:mat1}. The analyses were run with the quasi-static implicit solver of the commercial software Abaqus.
	
	 \begin{table}
		\caption{Mechanical parameters of the poro-hyperelastic model} \label{tab:mat1}
		\begin{tabular}{l c c}
			Ogden's parameters & {Brain and CH} $ ^1 $ & {Gelatine (10\% w/w)} $ ^2 $ \\
			\hline
			$ \mu $ (Pa) & $ 0.52 \cdot 10^{3} $ & $ 6.21 \cdot 10^{3} $ \\
			$ \alpha $ & -4.4 & 2.64 \\
			$ D $ $ \mathrm{Pa^{-1}} $ & $ 1.3 \cdot 10^{-3} $ & $ 69 \cdot 10^{-6}$ \\
			\hline
			Hydraulic conductivity $ k $ $ \mathrm{m\,s^{-1}} $ & $ 1.57 \cdot 10^{-9} $ & $ 1.25 \cdot 10^{-6} $ \\
			Fluid specific weight $ \gamma_F $ ($ \mathrm{kN\,m^{-3}} $) & 9741 & 9741 \\
			Initial void ratio $ e $ (\%) & 20 & 90 \\
			\hline
			Intrinsic toughness $ \Gamma_o $ $ \mathrm{J\,m^{-2}} $ & 0.1 & 1.1 \\
		\end{tabular}
	\\$ ^1 $ \citet{Forte2017} 
	\\$ ^2 $ \citet{Forte2015a} 
	\end{table}

Firstly, the case of an edge-crack subjected to far-field loading, by means of applied displacements $ U $ in the direction perpendicular to the crack, is considered. Then, we have studied the insertion of wires with diameter $ d_w=0.125-1\,\mathrm{mm} $ that are pushed into the crack for its full length. Although frictional effects are not considered in the model, a small coefficient of Coulomb's friction was introduced in the analyses because we have found that it helped to achieve numerical convergence of the contact algorithm. From the deformed coordinates $ y_1,y_2 $, the radius of the blunted crack can be expressed through the radius $ \varrho $ of a circle fitting the profile within a distance equal to $ 10^{-3} c $ from the crack tip. Plots of the deformed crack when $ J=\Gamma_o $ are shown in Fig.\ref{fig:wire}b, suggesting that the crack blunting with wires of various diameters is almost equivalent to the edge-crack subjected to remote loading. The transition to constrained blunting seems to occur when $ d_w=0.125 \mathrm{mm} $, which displays a markedly different trend. Interestingly, such a value is a good approximation of the characteristic length $ 2\varrho_o =2\Gamma_o/E $, which for the soft tissues of our study is in the order of $ 1.3 \cdot10^{-4}\mathrm{m} $. By plotting the critical crack tip radius $ \varrho $ against the wire diameter $ d_w $ Fig.(\ref{fig:wire}c), we notice that it is approximately equal to $ {\varrho_o} $ when $ d_w \ge 2{\varrho_o} $. Below this limit, we hypothesise that the tip radius scales with the wire diameter (hence the slope 1/2 shown in the plots). In conclusion, we can assume that, in the materials under consideration, steady state cutting is equivalent to a problem of crack propagation when the wire diameter is $ d_w \ge 0.13 \mathrm{mm} $.

\subsubsection{Fracture in the biphasic medium}

\begin{figure*}[h!]
	\includegraphics[scale=1]{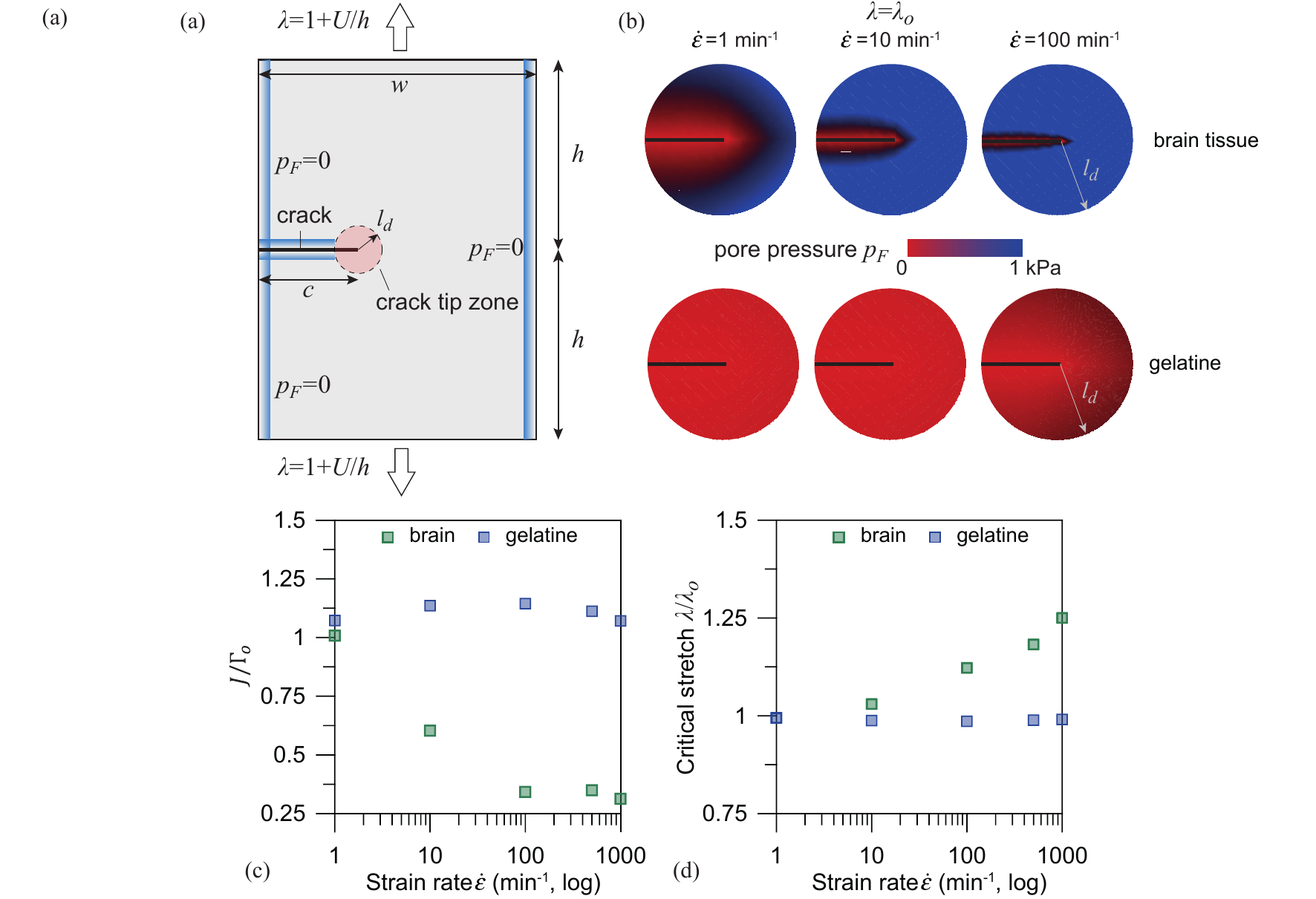}
	\caption{(a) Model of the edge-cracked sample for the analyses of poroelastic fracture. Blue edges are those with drained boundary conditions. (b) Contours of the pore pressure $ p _F $ at constant stretch $ \lambda=\lambda_o $ for different strain rates, in the brain tissue and the gelatine studied by \citet{Forte2015a}.(c) Strain energy per unit area in the crack-tip region, normalised by the fracture toughness $ \Gamma_o $, and (d) applied stretch, normalised by the critical stretch in quasi-static conditions $ \lambda_o $. Both are plotted as a function of the strain rate $ \dot \varepsilon $ (logarithmic plot). \textcolor{red}{2-col figure}}
	\label{fig:poro}
\end{figure*}

Retaining the assumption of autonomous crack propagation, we can study the rate-dependent fracture in an equivalent edge-cracked model with applied remote displacements. The geometry is illustrated in Fig.\ref{fig:poro}a: it consists of a large rectangular sample of height $ 2h=50\mathrm{mm} $ and width $ w=20\mathrm{mm} $, containing an edge-crack of length $ c=1\mathrm{mm}$. Normal displacements are applied to the top and bottom boundaries such that the strain rate is constant, that is $ U  = \left[ {{\mathrm{exp}({\dot \varepsilon t}}) - 1} \right]{h} $, where $ \varepsilon=\mathrm{ln}[(h+U)/h] $ is the true strain in the direction normal to the crack line. The stretch ratio is defined by $ \lambda=1+U/h $. Two materials, the brain tissue and the gelatine studied by \citet{Forte2015a}, are described with the poro-hyperelastic model presented in Sect.\ref{subsec:model}. The properties are summarised in Table \ref{tab:mat1}. Notice that the hydraulic conductivity $ k $ is employed in place of the permeability $ \kappa $, to which is related by means of $ k=\kappa {\gamma_F}/{\eta_F} $. The finite element mesh is built with four-node quadrilateral plane strain hybrid elements with additional degrees of freedom for the pore pressure $ p_F $. Boundary conditions are specified in terms of displacements (top and bottom forces are prevented from lateral motion), and in addition on the pore pressure degree of freedom. A condition of draining, enforced by setting the pore pressure equal to zero, is specified for the vertical free edges and the edge-crack surfaces in contact with atmospheric pressure (Fig.\ref{fig:poro}a). The reference porosity $ n_0 $ needs to be specified as initial condition through the void ratio $ e $. The analyses were run with the implicit solver of the commercial software Abaqus. A transient fluid-stress diffusion analysis is required to simulate fluid flow through the porous material, where the accuracy of the solution is governed by the maximum pore pressure change allowed in an increment. Different values have been considered for the best compromise between accuracy and efficiency.

The main purpose of the analyses is to understand how fluid draining in the crack-tip region affects the onset of crack propagation. In other terms, we are considering the effect of dissipation and of the loading rate on the crack driving energy, whereas the fracture toughness is assumed equal to $ \Gamma_o $. The critical condition is then defined by
\begin{equation}\label{eq:Gc-rate}
	J(\dot{\varepsilon})=\Gamma_o
\end{equation}
where $ J $ denotes the \textit{J}-integral, which in fracture mechanics is used to compute the driving force for crack growth. Due to coupling between deformation of the solid network and fluid diffusion, the \textit{J}-integral is path-dependent since it includes poroelastic dissipation, unless a vanishing small contour surrounding the crack tip is considered \citep{Wang2012b}. In this work we compute the \textit{J}-integral in the biphasic material according to \citep{Shih1986}
\begin{equation}\label{eq:Jint-1}
	J=\int _C \left ( \Psi n_1 - \frac{\partial u_i}{\partial x_1} \sigma'_\mathit{{ij}} n_{j}\right )\mathrm{d}s,
\end{equation}
where $ \Psi $ is the strain energy density and $ n_j $ is the unit vector normal to a contour $ C $ enclosing the crack tip (Fig.\ref{fig:wire}a). The results presented below are obtained considering a contour that surrounds the crack-tip region of radius $ l_d $.

A preliminary analysis on the elastic material is employed to investigate the quasi-static situation. In such a case, the critical condition evaluated through Eqs.(\ref{eq:Gc-rate})-(\ref{eq:Jint-1}) provides the stretch $ \lambda_o $, corresponding to the onset of crack propagation. Results in the poroelastic materials are illustrated in Fig.\ref{fig:poro}b, where we show the contours of the fluid pressure $ p_F $, for three different strain rates $ \dot \varepsilon $, when $ \lambda=\lambda_o $. The enlarged region is the crack-tip dissipative zone, whose radius, from Eq.(\ref{eq:R}), is approximated to $ l_d \sim 10^{-5} \mathrm{m} $. The red areas correspond to the drained or relaxed condition ($ p_F=0 $) whereas the blue regions are affected by fluid flowing in the pores. It can be seen that, independently from the rate, the greater permeability of gelatins allows for a rapid draining of the whole crack-tip region. On the contrary, it appears that fluid takes a longer time to drain the same area in the brain tissue, where permeability is much lower. Since fluid draining is a dissipative process, it is reasonable to assume that crack propagation is affected by the phenomenon, at least in the brain tissue. Keeping in mind the limitations in the use of the \textit{J}-integral, in Fig.\ref{fig:poro}c we present the normalised energy at constant stretch $ \lambda_o $ for different strain rates. The observed behaviour can be better comprehended by plotting the normalised stretch when $ J=\Gamma_o $, Fig.\ref{fig:poro}d. As expected, no difference with respect to the elastic quasi-static situation emerges in the gelatine, which therefore behaves as an elastic relaxed material. The situation looks different in the brain tissue, where both the strain energy and the critical stretch are affected by rate. Notice that we cannot consider these stretches as the real ultimate stretches of the material; nevertheless, the results shown in Fig.\ref{fig:poro}d suggest a toughening effect due to fluid draining in the brain tissue.
	
	\section{Discussion}
	\label{sec:disc}
	
	The phenomenon of poroelastic relaxation is characterised by a long-range motion of interstitial fluid in the solid network \citep{Wang2012a}. The characteristic time of relaxation depends on the length of observation, and this implies that samples with macroscopic sizes, such as those employed in the experiments, require relatively long times compared, for instance, to viscoelastic relaxation \citep{Hu2012}. However, with respect to the presence of a crack the situation is completely different. The topic of fracture in the brain tissue is not well documented; for this reason, we have developed our model based on the observations in hydrogels, which can be considered as benchmark examples of soft porous materials. 
	
	In the crack tip zone of soft porous tissues, high uniaxial tensile stresses trigger a take-up of fluid from the far-field region \citep{Long2015}. Since these high stress gradients are confined to a small region, whose extension in the soft tissues under investigation is in the order of $ 10^{-5}\,\mathrm{m} $, fluid draining of the crack tip zone is a relatively quick phenomenon. Evidently, it might happen that viscoelastic relaxation occurs at the same time, but this aspect has not been considered in the present work. The key observation is that material relaxation is relevant with respect to rate-dependent fracture in relation to the crack-tip dissipative region. In our analysis, we have explained this fact with separation of length scales, through which we can neglect the dissipative phenomena in the bulk; furthermore, we have assumed that the rate-independent toughness threshold is originated within the process zone, whose extension is in the scale of nanometers. There are models that explained fluid-related toughening based on the fluid draining in the process zone \citep{Forte2015a}. Interestingly, the model by \citet{Forte2015a} was proposed to explain the toughening of gelatins, which cannot be captured by the analyses presented in this work (Figs.\ref{fig:poro}b-d). 
	
		\begin{figure}[h!]
		\includegraphics[scale=1]{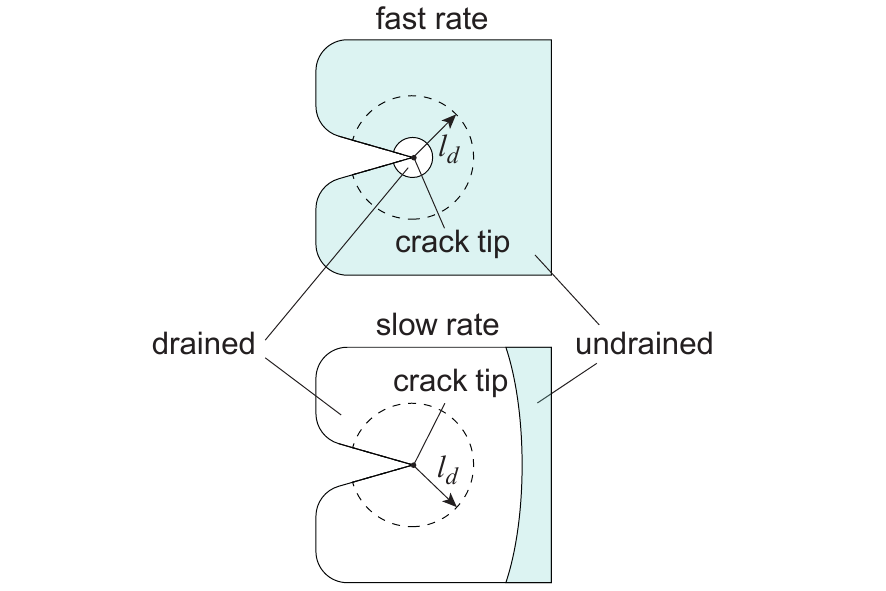}
		\caption{Sketch of the crack-tip dissipative region and influence of fluid draining at fast and slow strain rates $ \dot{\varepsilon} $. \textcolor{red}{1-col figure}}
		\label{fig:tip}
	    \end{figure}
	
	With specific consideration of fluid-related effects, the mechanism of crack-tip draining is illustrated in Fig.\ref{fig:tip}. Initially the solid is saturated and there is only an infinitesimal zone close to the tip of the existing crack where the fluid pressure is zero; as time goes by, this region increases in extent until the whole crack-tip region is drained. According to Biot’s theory, the pressure-driven fluid flow is a diffusive process. The time of poroelastic relaxation can be defined as the time needed to drain an area of radius $ l_d $ close to the crack-tip \citep{Hui2013}
	\begin{equation}\label{eq:diffusion}
		t_d={l_d^2}/{D_F}
	\end{equation}
	where  $ D_F $ is the diffusion coefficient, which depends on the permeability, the fluid viscosity and the elastic properties of the solid. To a first approximation, in linear poroelasticity and plane strain conditions we have $ D_F= {2\mu(1-\nu)\kappa}/{(1-2\nu)\eta_F} $, where $ \mu $ and $ \nu $ are, respectively, the shear modulus and Poisson's ratio of the solid material \citep{Hui2013}. Ideally, we can distinguish two limit situations:
	\begin{itemize}
		\item fast strain rates or reduced permeability: fluid diffusion is too slow to be effective and the crack tip region is saturated at the instant corresponding to the onset of propagation. With respect to fracture, the material behaves as a soft incompressible solid;
		\item slow strain rates or high permeability: the draining process is fast, therefore a large drained region surrounds the crack-tip zone. Here we observe an extensive process of draining but this becomes ineffective with respect to fracture. The material behaves as a soft compressible solid.
	\end{itemize}

	Intermediate situations are those in which fluid drains the crack-tip region in a time range comparable to that leading to crack propagation, hence causing a dissipative phenomenon that may produce an enhancement of the material toughness. This effect was correctly described by our model in the brain tissue. However, there are some limits in the procedure adopted, which are discussed below, that leave space for further work on this topic.
	
	In the procedure through which we have derived the toughness of brain tissue from the experimental wire cutting force, we have assumed that the slope of the steady state cutting force $ F_{SS} $ versus the wire diameter $ d_w $ is the same for different velocities (Fig.\ref{fig:exp}d). However, wire cutting analyses on biomimicking gelatins by \citet{Forte2015a} have shown that the slope of the interpolating function increases with the insertion velocity, although it tends to become constant at lower velocities. Furthermore, the extrapolation procedure to infer the intrinsic toughness $ \Gamma_o $ from Eq.(\ref{eq:cutting}) was based on the assumption that $ v=1\,\mathrm{mm\,min^{-1}} $ is a reasonable speed for quasi-static conditions. To a first approximation, we might relate the quasi-static threshold to the process of fluid diffusion, which in turn depends on the permeability of the material. The employed value was derived from similar observations by \citet{Forte2015a} on gelatins. However, considering the large difference with the permeability of brain tissue, our assumption needs to be verified against further experimental observations. This point brings us to a central aspect in modelling fluid-related effects in soft tissues: the issue of accurately measuring and modelling permeability. For the sake of simplicity, we have adopted the hypothesis of material isotropy: however, while it seems to be a valid assumption for the elasticity of the brain tissue \citep{Budday2020a}, diffusion or permeability properties are remarkably anisotropic, in particular in white matter regions characterised by axonal structures. In addition, brain tissue permeability can be modified substantially under loading by swelling and additional coupling with the local tissue deformation \citep{Jamal2020}.
	
	The model that we have developed isolated the stage of crack propagation, leaving aside the whole process of contact and indentation that occurs in wire cutting. Ideally, we could have simulated the complete cutting process directly through the finite element model and use cohesive elements to simulate the process of propagation. This approach has successfully modelled needle penetration in soft elastic materials \citep{Oldfield2013, Terzano2020} and rate-dependent wire cutting of viscous food \citep{Goh2005, Skamniotis2020}. However, to calibrate the cohesive model for the brain tissue would require the characterisation of its frictional behaviour \citep{Casanova2014}, which is known to affect the fracture toughness of the material \citep{Duncan2020}. In addition, the role of fluid on the frictional contact between wire and tissue should also be considered \citep{Reale2017}, possibly implying some effect of lubrication which, at the present time, is unknown. The FE model provided meaningful results on poroelastic toughening with rate (Figs.\ref{fig:poro}c-d) but we have no means to establish a direct confrontation with experimental data. Experiments revealed a rate-dependent toughening in terms of the velocity of wire insertion (Figs.\ref{fig:exp}e), which in the steady state can be reasonably considered to coincide with the crack velocity. In the numerical model we have instead explored the effect of the strain rate on the onset of crack propagation, but we cannot establish an analytical relationship between the strain rate and the crack propagation velocity. It comes naturally to think that higher strain rates result in faster crack propagation, although this might hold only below a certain limit, as shown for instance in fracture tests on hydrogels \citep{Mayumi2016}.
	
	Finally, although our poro-hyperelastic model is able to couple large deformation and fluid flow, more advanced models considering the full coupling between the solvent diffusion and tissue swelling might be required, e.g. \citet{Hong2008a, Bouklas2015, Chen2020, Brighenti2020}. This would also allow us to implement a modified definition of the \textit{J}-integral proposed for swelling materials, which is path-independent and computes the transient energy energy release rate by separating the energy lost in diffusion from the energy available to drive crack growth \citep{Yang2006, Bouklas2015a}.
	
	\section{Conclusion}
	\label{sec:conc}
	
	Testing the fracture properties of super soft tissues through standard tensile specimens is a complex task. For this reason, wire cutting was here employed to analyse the influence of rate on the fracture energy of brain tissue. The experimental data show an evident increase of the cutting force with the rate of insertion, suggesting that some form of energy dissipation affects the cutting process. In this work, we speculate that the rate-dependent toughening is due to poroelastic dissipation in the vicinity of the crack that is propagated ahead of the wire. We have proposed a numerical model which considered the brain tissue as a biphasic material. Through finite elements analyses of an edge-cracked sample, subjected to remote loading with varying strain rate, we have shown how the process of fluid draining in the crack-tip region might affect the fracture toughness of the material. We can then summarise the main findings:
	\begin{itemize}
		\item the analysis of wire cutting experimental data suggests a power-law increase of fracture toughness with the rate of insertion, as already observed in biomimicking gelatins;
		\item we have identified a length scale which distinguishes the fracture process of cutting from crack propagation under remote loading. Specifically, below a certain wire diameter crack propagation becomes unstable and the shape of the crack is constrained. Interestingly, this was observed experimentally on hydrogels \citep{Baldi2012} and was motivated by the reduced stiffness of such materials. More correctly, we are able to say that it depends on the competition between the cost of creating new surfaces and the elastic strain energy of the material;
		\item the finite element analyses of the fracture process in the poroelastic material have confirmed the toughening effect with the rate of applied loading. According to our poro-hyperelastic model, such a contribution is chiefly controlled by the value of the intrinsic permeability of the material.
	\end{itemize}
	
	This work has purposely neglected the dissipative behaviour provided by viscoelasticity in order to focus on fluid-related effects. Future work will be dedicated to extend the proposed model to coupled viscoelasticity and fluid diffusion. In the context of fracture, accurate models should specifically target the rate-sensitivity of the process zone. By a computational point of view, cohesive models might still be the ideal candidates to include energy dissipation through a time-dependent cohesive law. Our view is that they should be developed on the ground of a micromechanical description of the disintegrating material ahead of the crack tip. In particular, we envisage that further research is needed to characterise the effect of water diffusion on mechanical deformation by a micromechanical point of view.
	
    \section{Acknowledgements}
    \label{sec:ack}
    A. E. Forte acknowledges the support received from the European Union’s Horizon 2020 research and innovation programme under the Marie Skłodowska-Curie grant agreement No 798244. The authors also acknowledge the financial support from EDEN2020 project funded by the European Union’s Horizon 2020 research and innovation programme under grant agreement No 688279. D. Dini would like to acknowledge the support received from the UKRI Engineering and Physical Sciences Research Council (EPSRC) via his Established Career Fellowship EP/N025954/1.
	
	\bibliographystyle{elsarticle-harv.bst}
	\bibliography{library}
	
\end{document}